# Spatial Autocorrelation Approaches to Testing Residuals from Least Squares Regression

Yanguang Chen

Department of Geography, College of Urban and Environmental Sciences, Peking University, 100871, Beijing, China. Email: chenyg@pku.edu.cn

**Abstract**: In statistics, the Durbin-Watson test is always employed to detect the presence of serial correlation of residuals from a least squares regression analysis. However, the Durbin-Watson statistic is only suitable for ordered time or spatial series. If the variables comprise cross-sectional data coming from spatial random sampling, the Durbin-Watson will be ineffectual because the value of Durbin-Watson's statistic depends on the sequences of data point arrangement. Based on the ideas from spatial autocorrelation, this paper presents two new statistics for testing serial correlation of residuals from least squares regression based on spatial samples. By analogy with the new form of Moran's index, an autocorrelation coefficient is defined with a standardized residual vector and a normalized spatial weight matrix. Then on the analogy of the Durbin-Watson statistic, a serial correlation index is constructed. As a case, the two statistics are applied to the spatial sample of 29 China's regions. These results show that the new spatial autocorrelation model can be used to test the serial correlation of residuals from regression analysis. In practice, the new statistics can make up for the deficiency of the Durbin-Watson test.

# 1 Introduction

The least squares regression can be employed to make models of real systems for revealing the relationships between causes and effects. The major aims of mathematical modeling lie in explanation and prediction, which are sometimes contradictory (Batty, 1991; Kac, 1969;

Fotheringham and O'Kelly, 1989). By means of regression modeling, we can explain the causes for an effect, or predict the effects with causes. The quality of a mathematical model depends on its structure. A model must simplify reality to the moment. As Longley (1999, page 605) pointed out: "In the most general terms, a 'model' can be defined as a 'simplification of reality', nothing more, nothing less." Both oversimplification (explanatory variables are incomplete) and undersimplification (explanatory variables are redundant) of reality will lead to trustless explanation and unfaithful prediction. The structural problems of a model can be reflected by residuals, that is, a series of errors between observed values and predicted values given by the model. A good model is supposed to yield a random series of residuals with no autocorrelation. The autocorrelation phenomena in the residual series suggest an inherent defect in the model.

One of approaches to residual autocorrelation analysis of linear regression models is the Durbin-Watson test. Durbin and Watson (1950, 1951, and 1971) once wrote a series of articles to develop a method of testing for serial correlation in a least squares regression. One of the fruits is the well-known Durbin-Watson's statistic, which is easy to understand, calculate, and explain. New test statistics have been derived from the standard Durbin-Watson assumptions, being uniformly most powerful against the alternative hypothesis that the errors stemming from the stationary first order Markov process (Sargan and Bhargava, 1983). However, Durbin-Watson's method has a significant limitation, that is, it cannot be applied to the regression analysis based on cross-sectional data, which is defined in a 2-dimension space. The Durbin-Watson formula is constructed with one-order time lag or one step space displacement. Therefore, it is only applicable to the least squares regressions based on ordered time series or spatial series, which is defined in a 1-dimension space. An ordered time series or spatial series has exclusive rank for observed data of a variable, thus the result of Durbin-Watson statistic is uniquely determined. However, an array of cross-sectional data has various arrangement orders for numerical data. Changing the rank of elements in an array will result in a different series of residuals, and thus lead to different values of the Durbin-Watson statistic. In particular, in geographical analysis, many least squares regressions are based on cross-sectional data from spatial random sampling. The Durbin-Watson test is often ineffective in the linear regression of spatial analysis.

A new way of testing the serial correlation of residuals from least squares regression based on cross-sectional data is to make use of spatial autocorrelation analysis. Actually, the core of the formula of Durbin-Watson's statistic is just a 1-dimensional autocorrelation coefficient. Using a weight function to replace the time-lag parameter or space-displacement parameter, the 1-dimension spatial or temporal autocorrelation model can be generalized to 2-dimensioan spatial autocorrelation model. There are two basic and important statistics for spatial autocorrelation. One is Moran's index (Moran, 1950), and the other, Geary's coefficient (Geary, 1954). The former is presented by generalizing Pearson's correlation coefficient (Appendix 1), while the latter is constructed by analogy with Durbin-Watson's statistic (Appendix 2). Moran's index can be re-expressed in a very simple equation using standardized vector and normalized spatial weight matrix (Chen, 2013). Based on the new mathematical expression of Moran's index, a relatively precise formula of the residual autocorrelation can be defined. Further, by analogy with Geary's coefficient, an approximate expression of the residual autocorrelation can be put forward. The rest of this paper is organized as follows. Section 2 gives the basic expressions of 2-dimensional spatial autocorrelation of residuals, and the concept of residual autocorrelation scatterplot is proposed; Section 3 displays a set of case analysis to show how to use the methods presented in this work to make a testing for serial correlation; In Section 4, the 2-dimensional spatial autocorrelation measurement of residuals is generalized and developed, and the deficiency of these measurements is discussed. Finally, the paper is concluded with summarizing the highlights of this study.

## 2 Models and methods

### 2.1 A deficiency in the Durbin-Watson test

The aim of this study is to solve the problem of residuals test of regression analysis based on spatial data from a new angle of view. It is necessary to explain the general linear regression model and its residuals. Suppose there are *m* variables ($j$=1, 2,…, *m*) and n spatial elements in region ($i$=1, 2,…, *n*). In this instance, the sample size is *n*. The multivariable linear regression equation can be expressed as

$$y_i = a + \sum_{j=1}^{m} b_j x_{ij} + \varepsilon_i, \tag{1}$$

where $x_i$ denotes independent variables (input variables, explanatory variables, arguments), $y_i$ represents a dependent variables (output variable, explained variable, function), $a$ refers to a constant (intercept), $b_j$ to regression coefficients (slopes), and $\varepsilon_j$ to residuals (errors). The residuals are supposed to be a white noise series and must satisfy the following conditions

$$\varepsilon_i \sim WN(0, \sigma^2), \tag{2}$$

where "WN" means "white noise". That is to say, the average value of the residual series must be 0, and its limited variance is a constant $\sigma^2$. If and only if the residual series is a white noise, it will imply that the errors between the observed values and predicted values of the regression model, equation (1), come from the random disturbance outside the model, otherwise, it will mean that the errors result from the internal structure of the model itself. One of approaches to judging whether or not the residual series is a white noise is the well-known Durbin-Watson statistic (ab. DW), which is defined as

$$DW = \frac{\sum_{i=2}^{n}(\varepsilon_i - \varepsilon_{i-1})^2}{\sum_{i=1}^{n}\varepsilon_i^2} = \frac{\sum_{i=1}^{n-1}(\Delta\varepsilon_i)^2}{\sum_{i=1}^{n}\varepsilon_i^2} = 2(1-\rho), \tag{3}$$

where $\Delta\varepsilon_i = \varepsilon_i - \varepsilon_{i-1}$, and

$$\rho = \frac{\sum_{i=2}^{n}\varepsilon_i \varepsilon_{i-1}}{\sum_{i=1}^{n}\varepsilon_i^2} \tag{4}$$

denotes the autocorrelation coefficient of the residual series. In equations (3) and (4), the difference of $i$ indicates a one-order time lag ($k=\Delta i =1$) or a one-step space displacement ($r=\Delta i=1$). Because the coefficient $\rho$ comes between -1 and 1 (i.e., $-1 \leq \rho \leq 1$), the $DW$ values vary from 0 to 4 (i.e., $0 \leq DW \leq 4$). If the residuals have no serial correlation, then $\rho=0$, and thus $DW=2$. This suggests that if the DW statistic is close to 2, the residual series can be regarded as free of autocorrelation at a certain significance level (say, $\alpha=0.05$).

However, the Durbin-Watson test is only applicable to the serial correlation of residuals from the least squares regression based on ordered times series, say, the US level of urbanization from 1790 to 2010, or spatial series, say, the average urban population density of the rings from the center of a city to its exurbs (Chen, 2008). If we make a regression analysis using cross-sectional data coming from spatial random sample, the Durbin-Watson method will be ineffective (Table 1). As indicated by equation (3), the *DW* value is calculated with residuals and the sum of squares of the differences of residuals, but the difference sum of squares depends on the arrangement of elements in a random sample. For the cross-sectional data, the elements can be randomly arranged in spreadsheet. The results from different data arrangement will be different from one another. For example, suppose that for the set of elements [A, B, C], the corresponding array is [1, 2, 3]. Thus the vector of difference is [1, 1], and the sum of squares of the differences is 2. If this is an ordered temporal or spatial set, the order of A, B, and C cannot be changed. However, for a spatial random sampling, the arrangement of the elements is arbitrary. If the elements is permuted and the result is [B, A, C], the corresponding array will change to [2, 1, 3]. Then the difference vector is [-1, 2], and the sum of square of the differences is 5. This suggests that, for a spatial random sample, the *DW* value is not certain. It depends on the arrangement of the elements which are measured for sampling.

**Table 1 The sphere of application of the Durbin-Watson test method for serial correlation**

| Object | Data set | DW statistic | Cause |
| --- | --- | --- | --- |
| **Ordered time series** | Longitudinal data sequence | Effective | The differences of residuals are determinate |
| **Ordered spatial series** | Spatial data sequence | Effective | The differences of residuals are determinate |
| **Random spatial series** | Cross-sectional data | Ineffective | The differences of residuals depend on element arrangement |

## 2.2 A new approach to test serial correlation

An effective approach to solving this problem is to make use of spatial autocorrelation. Moran's index is in fact a spatial autocorrelation coefficient. The mathematical expression of Moran's index has been simplified by means of standardized vectors and a unitized matrix (Chen, 2013). Using the normalized form of the formula of Moran's index, we can construct new statistics of testing for the serial correlation of the residuals from the least squares regression based on spatial random samples. The series of residuals in equation (1) can be standardized with the formula

$$e_i = \frac{\varepsilon_i}{\sigma}, \tag{5}$$

where $\sigma$ denotes the standard deviation of the residual series. If the spatial distance matrix of the random sampling points has been obtained, we will have an *n*-by-*n* unitary spatial weights matrix (SWM) such as

$$W = [w_{ij}]_{n \times n}. \tag{6}$$

The three properties of this matrix are as follows: (1) Symmetry, i.e., $w_{ij}=w_{ji}$; (2) Zero diagonal elements, i.e., $|w_{ii}|=0$, which implies that the entries in the diagonal are all 0; (3) Unitary condition, that is

$$\sum_{i=1}^{n}\sum_{j=1}^{n} w_{ij} = 1. \tag{7}$$

Thus the spatial autocorrelation coefficient of the residuals can be computed by the following formula

$$I = e^{\mathrm{T}} W e, \tag{8}$$

where *I* denotes spatial autocorrelation index (SAI) of residuals, which is equivalent in mathematical expression to Moran's index. The index ranges between -1 and 1 (i.e., $-1 \le I \le 1$). If the residuals have no serial correlation, we will have $I=0$. By analogy with the Durbin-Watson statistic expressed with equation (3), the residual correlation index (RCI) of least squares regression can be defined as

$$S = 2(1 - I), \tag{9}$$

where *S* indicates RCI. Apparently, the *S* value comes between 0 and 4 (i.e., $0 \le S \le 4$). If the *S* value is close to 2, we will reach a conclusion that the residuals have no spatial autocorrelation according to a certain significance level.

## 2.3 Residuals correlation scatterplot

On the analogy of the normalized Moran's scatterplot, a residual autocorrelation scatterplot can constructed for serial correlation analysis. Because $e^{\mathrm{T}}e=n$, equation (8) can be expressed as

$$e^{\mathrm{T}} e I = e^{\mathrm{T}} (nW) e, \tag{10}$$

This suggest the precondition that equation (10) comes into existence is as follows

$$nWe = Ie. \tag{11}$$

That is to say, from equation (11) it follows equation (8). On the other hand, equation (8) multiplied left by $e$ on both sides of the equal sign yields

$$Ie = ee^{\mathrm{T}}We. \tag{12}$$

Based on equation (11), a random variable based on observed values can be defined in the form

$$y = e^{\mathrm{T}}eWe = nWe. \tag{14}$$

Based on equation (12), a trend variable based on predicted values can be defined as below

$$\hat{y} = ee^{\mathrm{T}}We = Ie. \tag{15}$$

Then, using $e$ as $x$-axis and $y$ as well as $\hat{y}$ as $y$-axis, we can make a serial correlation scatterplot. In the plot, the relationships between $e$ and $y$ give the scattered points, and relationships between $e$ and $\hat{y}$ yields the trend line. The slope of the trend line is equal to the SAI value.

## 2.4 Developed and alternative mathematical forms

An effective definition of RCI and the related analytical process for random serial correlation have been proposed above. In fact, the mathematical expressions and calculation methods are diverse. The RCI can be given in two forms: one is relatively precise form based on Moran's index, the other is an approximate form based on Geary's coefficient. On the other hand, in theory, the RCI is expressed in the form based on population (universe); while in practice, it always takes the form based on sample. No matter what form it is, a spatial contiguity matrix (SCM) must be constructed (Chen, 2012). Suppose there are $n$ elements in a geographic region. A SCM can be expressed as

$$V = [v_{ij}]_{n \times n} = \begin{bmatrix} v_{11} & v_{12} & \cdots & v_{1n} \\ v_{21} & v_{22} & \cdots & v_{2n} \\ \vdots & \vdots & \ddots & \vdots \\ v_{n1} & v_{n2} & \cdots & v_{nn} \end{bmatrix}, \tag{16}$$

where $V$ denotes the SCM, $v_{ij}$ is a measure used to compare and judge the degree of nearness or the contiguous relationships between location $i$ and location $j$ ($i, j=1,2,\ldots,n$). For the elements on the diagonal, they are zeros, otherwise they must be turned onto zero (i.e.m for $i=j$, $v_{ii} \equiv 0$). A sum of SCM entries can be defined as

$$T = \sum_{i=1}^{n}\sum_{j=1}^{n} v_{ij} . \tag{17}$$

The SCM can be converted into SWM by the following formula:

$$w_{ij} = \frac{v_{ij}}{T} = v_{ij} / \sum_{i=1}^{n}\sum_{j=1}^{n} v_{ij} . \tag{18}$$

Based on population, equation (8) can be developed in detail to yield an expression similar to the formula of Moran's index, that is

$$I_{\mathrm{P}} = \frac{\varepsilon^{\mathrm{T}}(nW)\varepsilon}{\varepsilon^{\mathrm{T}}\varepsilon} = \frac{n\sum_{i=1}^{n}\sum_{j=1}^{n} v_{ij}(\varepsilon_i - \mu)(\varepsilon_j - \mu)}{T\sum_{i=1}^{n}(\varepsilon_i - \mu)^2} = \frac{n\sum_{i=1}^{n}\sum_{j=1}^{n} w_{ij}\varepsilon_i\varepsilon_j}{\sum_{i=1}^{n}\varepsilon_i^2} . \tag{19}$$

where $\mu=0$ denotes the mean of residuals. Thus the RCI based on population can be expressed as

$$S_{\mathrm{p}} = 2(1 - I_{\mathrm{p}}) . \tag{20}$$

Based on samples, equation (19) can be revised as

$$I_{\mathrm{s}} = \frac{\varepsilon^{\mathrm{T}}[(n-1)W]\varepsilon}{\varepsilon^{\mathrm{T}}\varepsilon} = \frac{(n-1)\sum_{i=1}^{n}\sum_{j=1}^{n} w_{ij}\varepsilon_i\varepsilon_j}{\sum_{i=1}^{n}\varepsilon_i^2} , \tag{21}$$

Accordingly, the RCI based on samples can be expressed as

$$S_{\mathrm{s}} = 2(1 - I_{\mathrm{s}}) . \tag{22}$$

As indicated above, Geary's coefficient is presented by analogy with the Durbin-Watson statistic. Now, on the analogy of the formula of Geary's coefficient, we can define serial autocorrelation index in the following form

$$C = \frac{(n-1)\sum_{i=1}^{n}\sum_{j=1}^{n} v_{ij}(\varepsilon_i - \varepsilon_j)^2}{2\sum_{i=1}^{n}\sum_{j=1}^{n} v_{ij}\sum_{i=1}^{n}(\varepsilon_i - \mu)^2} = \frac{(n-1)\sum_{i=1}^{n}\sum_{j=1}^{n} w_{ij}(\varepsilon_i - \varepsilon_j)^2}{2\sum_{i=1}^{n}\varepsilon_i^2} . \tag{23}$$

Thus an approximate residual correlation index (ARCI) can be defined as

$$S_a = 2C = \frac{(n-1)\sum_{i=1}^{n}\sum_{j=1}^{n} w_{ij}(\varepsilon_i - \varepsilon_j)^2}{\sum_{i=1}^{n}\varepsilon_i^2} . \tag{24}$$

Both the RCI and ARCI can be termed *spatial Durban-Watson* (SDW) statistics. In theory, we have

$$S_a = 2(1 - I_s) = S_s. \tag{25}$$

However, in practice, we have

$$S_a \approx 2(1 - I_s) = S_s. \tag{26}$$

This can be demonstrated by means of mathematical transformation, that is

$$S = 2C = \frac{n\sum_{i=1}^{n}\sum_{j=1}^{n} w_{ij}(\varepsilon_i - \varepsilon_j)^2}{\sum_{i=1}^{n}\varepsilon_i^2} = \frac{2n\sum_{i=1}^{n}\sum_{j=1}^{n} w_{ij}(\varepsilon_i^2 - \varepsilon_i\varepsilon_j)}{\sum_{i=1}^{n}\varepsilon_i^2}$$

$$= 2\left[\frac{\sum_{i=1}^{n}\sum_{j=1}^{n} w_{ij}\varepsilon_i^2}{\frac{1}{n}\sum_{i=1}^{n}\varepsilon_i^2} - \frac{n\sum_{i=1}^{n}\sum_{j=1}^{n} w_{ij}\varepsilon_i\varepsilon_j}{\sum_{i=1}^{n}\varepsilon_i^2}\right] \tag{27}$$

$$\approx 2[1 - I_p]$$

In fact, the Durbin-Watson statistic is an approximate measure rather than an exact measure for serial correlation test. ARCI is more similar to the DW index than RCI. Comprising equation (24) with equation (3) shows that there is a clear analogy between the Durbin-Watson statistic and the ARCI. The difference rests with that the one-order time lag in equation (3) is replaced by a spatial weight function in equation (24). For an even distribution of *n* elements and *n* is very large, we will have a weight $w_{ij} \to 1/n \approx 1/(n-1)$. The spatial difference $\varepsilon_i - \varepsilon_j$ bears an analogy with the temporal difference $\varepsilon_i - \varepsilon_{i-1}$. This suggests the Durbin-Watson formula defined by equation (3) and the ARCI defined by equation (24) are mathematically isomorphic to each other.

## 3 Application cases

### 3.1 Study area, problems, and analytical process

The method of spatial serial autocorrelation analysis can be applied to the least squares regression of the relationships between urbanization and economic development. The study area is China, which includes 31 provinces, autonomous regions, and municipalities directly under the Central Government of China. Two variables are employed to make the regression analysis: one is the *level of urbanization*, and the other, per capita *gross regional product* (GRP). The level of urbanization refers to the proportion of urban population to total population in a region. The statistical data of the urbanization levels and per capita GRP (2000-2013) are available from the website of National Bureau of Statistics (NBS) of the People's Republic of China (http://www.stats.gov.cn/tjsj/ndsj/). In order to implement the spatial serial correlation test, we need a spatial contiguity matrix. The matrix can be generated with the distances by train between any two capital cities of regions. The railroad distance matrix can be found in many Chinese road atlases. Because the cities of Haikou and Lhasa are not connected to the network of Chinese cities by railway from 2000 to 2013, only 29 regions and their capital cities are taken into consideration, and thus the size of each spatial sample is $n$=29 (Table 1). The datasets of urbanization level, per capita GRP, and railway distance are attached (S1). For the sample analysis, the number, $n$, should be replaced by total degree of freedom, $n$-1(Chen, 2013).

The analytical process consists of two operations: the first is regression analysis of the levels of urbanization and economic development, which yields a series of residuals; the second is the serial correlation test of residuals, which is based on the spatial autocorrelation analysis. Using per capita GRP as an independent variable and the level of urbanization as a dependent variable, we can make a regression analysis easily. The regression results include residuals ($\varepsilon_i$) and standardized residuals ($e_i$). Then, the three-step calculation method, which is designed for computing Moran's index (Chen, 2013), can be utilized to calculate the SAI, from which it follows RCI. Suppose that the residual vector $\varepsilon$ has been turned into the standardized vector $e$ using equation (5). This calculation is always fulfilled in the process of regression analysis by mathematical software or statistical software. Based on the vector of standardized residuals $e$ and the spatial weight matrix, $W$, we can get the RCI values through three steps as follows. **Step 1: calculate the normalized SWM**. The railway distance matrix can be turned into a spatial contiguity matrix with a weight function such as $v_{ij}$=1/$r_{ij}$, where $r_{ij}$ refers to the railway distance between city $i$ and city $j$, and $v_{ij}$ to spatial contiguity of the two cities (Chen, 2012). **Step 2: compute SAI**. In terms of equation (8), the SWM is first left multiplied by the transposition of $e$, and then the product of $e^T$ and $W$ is right multiplied by $e$. The final product of the continued multiplication is the SAI value. **Step 3: compute RCI**. It is very easy to calculate the RCI value by using equation (9).

## 3.2 Testing for serial correlation of linear regression analyses

The correlation between the level of urbanization and the level of economic development is one of current hot topics in China. The linear regression analysis can be employed to research the relationships between urbanization and economic development. Using the per capita GRD indicative of the economic development level as an argument and the proportion of urban population indicative of the level of urbanization as a response variable, we can build a simple linear regression model. Now, take the datasets of the year 2012 as an example. Suppose that the 29 Chinese regions are arranged in conventional order, which is in fact an official order. By using the least squares calculation, we will have a liner model is as below:

$$L_i = a + bG_i + \varepsilon_i = 26.1393 + 0.0006388 G_i + \varepsilon_i.$$

The goodness of fit is about $R^2=0.8944$ (Figure 1).

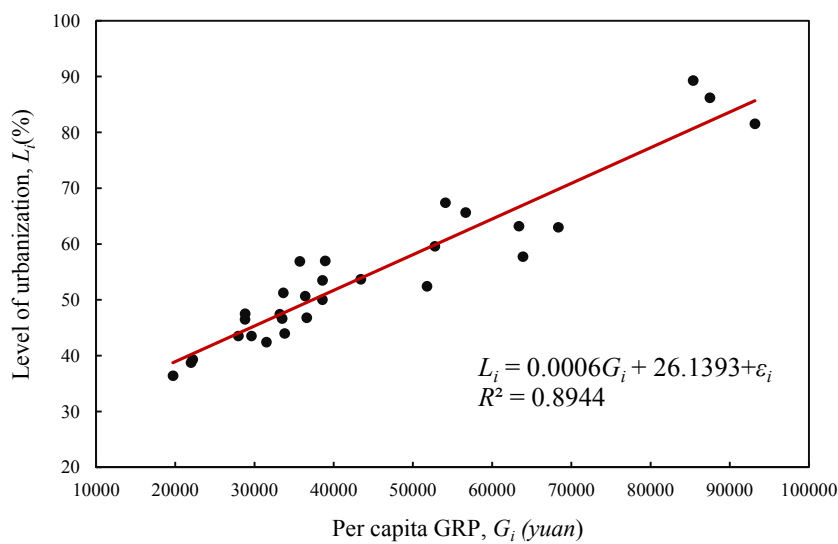

**Figure 1 The regression model of the linear relationship between urbanization and economic development of the 29 Chinese regions (2012)**

In order to appraise the model, a test for serial correlation of residuals must be performed. Based on the standardized value of the residuals $\varepsilon_i$, the Durbin-Watson statistic can be obtained using equation (3), and the result is about $DW=2.2463$. Then, by means of equations (8) and (9) and the abovementioned three-step method of calculation, we can compute the SAI and RCI, and the results are $SAI=-0.0915$ and $RCI=2.1830$. Note that the sample size $n$ is substituted with degree of freedom $n$-1. The basic process and main results of calculation are attached (S2). However, if we rearrange the elements of the spatial sample, the RCI value will not change, but the DW value will be different. For example, arranging the 29 regions in alphabetical order, we will have $DW$=1.9071 and $RCI$=2.1830. The RCI value is constant, but the DW value depends on the arrangement order of regions (Table 2). The corresponding computation process and results are attached (S3). Using $e$ as the $x$-axis, and $ee^T We$ and $(n-1)We$ as the $y$-axis, we can draw a normalized autocorrelation scatterplot of residuals as follows (Figure 2). The slope of the trendline is just equal to the SAI value, -0.0915.

Table 2 The datasets of per capita GRP, level of urbanization, and the standardized residuals from linear squares regression of 29 Chinese regions (2012)

| Arrangement in conventional order | | | | Arrangement in alphabetical order | | | |
| --- | --- | --- | --- | --- | --- | --- | --- |
| Region | per capita GRP | Level of urbanization | Residual | Region | per capita GRP | Level of urbanization | Residual |
| Beijing | 87475 | 86.20 | 0.9550 | Anhui | 28792 | 46.50 | 0.4496 |
| Tianjin | 93173 | 81.55 | -0.9400 | Beijing | 87475 | 86.20 | 0.9550 |
| Hebei | 36584 | 46.80 | -0.6196 | Chongqing | 38914 | 56.98 | 1.3671 |
| Shanxi | 33628 | 51.26 | 0.8315 | Fujian | 52763 | 59.60 | -0.0564 |
| Inner Mongolia | 63886 | 57.74 | -2.1058 | Gansu | 21978 | 38.75 | -0.3268 |
| Liaoning | 56649 | 65.65 | 0.7591 | Guangdong | 54095 | 67.40 | 1.5320 |
| Jilin | 43415 | 53.70 | -0.0399 | Guangxi | 27952 | 43.53 | -0.1066 |
| Heilongjiang | 35711 | 56.90 | 1.8165 | Guizhou | 19710 | 36.41 | -0.5305 |
| Shanghai | 85373 | 89.30 | 1.9705 | Hebei | 36584 | 46.80 | -0.6196 |
| Jiangsu | 68347 | 63.00 | -1.5549 | Heilongjiang | 35711 | 56.90 | 1.8165 |
| Zhejiang | 63374 | 63.20 | -0.7830 | Henan | 31499 | 42.43 | -0.8760 |
| Anhui | 28792 | 46.50 | 0.4496 | Hubei | 38572 | 53.50 | 0.6216 |
| Fujian | 52763 | 59.60 | -0.0564 | Hunan | 33480 | 46.65 | -0.2006 |
| Jiangxi | 28800 | 47.51 | 0.6793 | Inner Mongolia | 63886 | 57.74 | -2.1058 |
| Shandong | 51768 | 52.43 | -1.5500 | Jiangsu | 68347 | 63.00 | -1.5549 |
| Henan | 31499 | 42.43 | -0.8760 | Jiangxi | 28800 | 47.51 | 0.6793 |
| Hubei | 38572 | 53.50 | 0.6216 | Jilin | 43415 | 53.70 | -0.0399 |
| Hunan | 33480 | 46.65 | -0.2006 | Liaoning | 56649 | 65.65 | 0.7591 |
| Guangdong | 54095 | 67.40 | 1.5320 | Ningxia | 36394 | 50.67 | 0.2927 |
| Guangxi | 27952 | 43.53 | -0.1066 | Qinghai | 33181 | 47.44 | 0.0236 |
| Chongqing | 38914 | 56.98 | 1.3671 | Shaanxi | 38564 | 50.02 | -0.1726 |

| Sichuan | 29608 | 43.53 | -0.3484 | Shandong | 51768 | 52.43 | -1.5500 |
|---|---|---|---|---|---|---|---|
| Guizhou | 19710 | 36.41 | -0.5305 | Shanghai | 85373 | 89.30 | 1.9705 |
| Yunnan | 22195 | 39.31 | -0.2305 | Shanxi | 33628 | 51.26 | 0.8315 |
| Shaanxi | 38564 | 50.02 | -0.1726 | Sichuan | 29608 | 43.53 | -0.3484 |
| Gansu | 21978 | 38.75 | -0.3268 | Tianjin | 93173 | 81.55 | -0.9400 |
| Qinghai | 33181 | 47.44 | 0.0236 | Xinjiang | 33796 | 43.98 | -0.8571 |
| Ningxia | 36394 | 50.67 | 0.2927 | Yunnan | 22195 | 39.31 | -0.2305 |
| Xinjiang | 33796 | 43.98 | -0.8571 | Zhejiang | 63374 | 63.20 | -0.7830 |
| DW statistic | | 2.2463 | | DW statistic | | 1.9071 | |
| RCI | | 2.1830 | | RCI | | 2.1830 | |
| ARCI | | 2.1435 | | ARCI | | 2.1435 | |

**Note**: The unit of the level of urbanization is percent (%), and the unit of GRP is *yuan* of Renminbi (RMB).

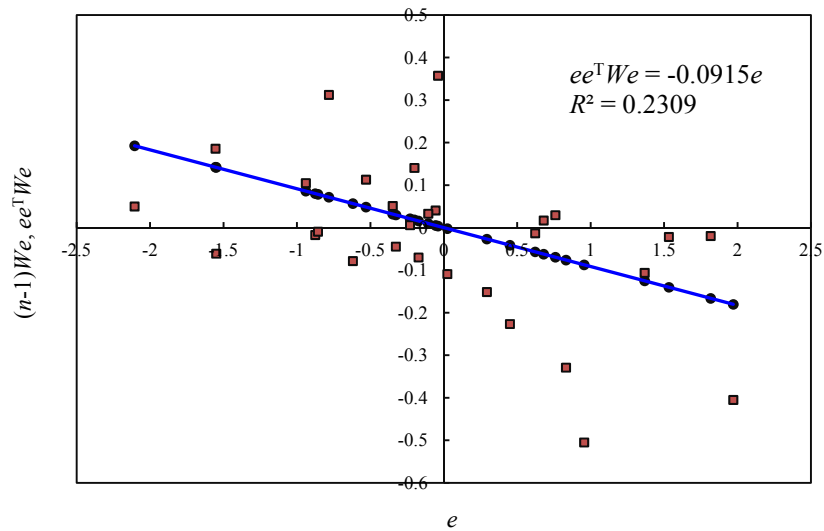

**Figure 2 The normalized scatterplot with a trendline of serial autocorrelation for the relationship between urbanization and economic development of the 29 Chinese regions (2012)**

The above method can be applied to the datasets of different years, from 2000 to 2012, and thus we will have 10 study cases (The statistical data of the level of urbanization are absent in the website of China's NBS). The weight functions are adopted to generate spatial contiguity matrixes. One is the inverse power function, $v_{ij}=1/r_{ij}$, and the other, a negative exponential function in the form $v_{ij}=\exp(-2r_{ij}/\bar{r})$, where $\bar{r}$ denotes the average distance. This study relies heavily on the inverse power function. The calculations based on the negative exponential function are for reference only. All the results are tabulated as follows (Table 3). Apparently, the RCI values are independent of arrangement order of the 29 regions, but they are dependent to a degree on the spatial weight function. However, the Durbin-Watson statistic values depend to a great extent on the arrangement order of sample data. For example, for the year of 2000, the Durbin-Watson statistic based on conventional order of regions is $DW$=1.5758, while the result based on the alphabetical order is $DW$=2.4939; for 2008, the two $DW$ values are 1.4310 and 1.9203, respectively. For the 29 regions, we have 29! sorts of permutations. This suggests that we can get about $8.8418*10^{30}$ DW values. Sometimes the numerical values of the Durbin-Watson statistic based on different permutations are considerably big. On the contrary, for given weight function, the RCI value is uniquely determined. Changing the weight function yields different RCI values. But generally speaking, there is no significant difference between the RCI values based on different weight functions. In short, the RCI value depends to some extent on weight functions but is independent of the permutation of elements.

Table 3 The Durbin-Watson statistics, RCI values, and ARCI values of residual series from linear squares regression of 29 Chinese regions (2000-2012)

| Year | Arrangement in conventional order | | | | | Arrangement in alphabetical order | | | | |
|---|---|---|---|---|---|---|---|---|---|---|
| | DW statistic | Power law based | | Exponential law based | | DW statistic | Power law based | | Exponential law based | |
| | | RCI | ARCI | RCI | ARCI | | RCI | ARCI | RCI | ARCI |
| **2000** | 1.5758 | 1.7576 | 1.7945 | 1.7493 | 1.7105 | 2.4939 | 1.7576 | 1.7945 | 1.7493 | 1.7105 |
| **2005** | 1.4621 | 1.7984 | 1.6745 | 1.8112 | 1.6243 | 1.9905 | 1.7984 | 1.6745 | 1.8112 | 1.6243 |
| **2006** | 1.5054 | 1.8135 | 1.6855 | 1.8352 | 1.6472 | 1.9345 | 1.8135 | 1.6855 | 1.8352 | 1.6472 |
| **2007** | 1.6049 | 1.8390 | 1.7364 | 1.8610 | 1.7029 | 1.9613 | 1.8390 | 1.7364 | 1.8610 | 1.7029 |
| **2008** | 1.4310 | 1.9045 | 1.7797 | 1.9168 | 1.7441 | 1.9203 | 1.9045 | 1.7797 | 1.9168 | 1.7441 |
| **2009** | 1.6044 | 1.9986 | 1.8986 | 1.9953 | 1.8635 | 1.8789 | 1.9986 | 1.8986 | 1.9953 | 1.8635 |
| **2010** | 1.8956 | 2.0418 | 1.9807 | 2.0240 | 1.9570 | 2.0448 | 2.0418 | 1.9807 | 2.0240 | 1.9570 |
| **2011** | 2.1046 | 2.1363 | 2.1068 | 2.0921 | 2.0565 | 1.9245 | 2.1363 | 2.1068 | 2.0921 | 2.0565 |
| **2012** | 2.2463 | 2.1830 | 2.1435 | 2.1329 | 2.0829 | 1.9071 | 2.1830 | 2.1435 | 2.1329 | 2.0829 |
| **2013** | 2.2524 | 2.2142 | 2.1755 | 2.1656 | 2.1055 | 1.8315 | 2.2142 | 2.1755 | 2.1656 | 2.1055 |

**Note**: "Power law based" means that the spatial contiguity matrix is generated with the inverse power function indicating of power-law decay. "Exponential law based" means that the contiguity matrix is yielded with a negative exponential function indicating exponential decay.

## 3.3 Testing for serial correlation of log-linear regression analyses

The relationship between the level of urbanization and that of economic development is impossibly a real linear relation. The reason for this is that the proportion of urban population has a clear *lower limit* (0) and a strict *upper limit* (1 or 100%). If a sample size is large enough, the distribution trend of the level of urbanization dependent on per capita GRP will be a curve instead of a straight line. Three equations can be employed to describe the relationship between urbanization and economic development. The first is the single logarithmic linear relation, which can be modeled with a logarithmic function (Dillinger, 1979; Zhou, 1989), the second is the double logarithmic linear relation, which can be modeled with a power function (Rao *et al*, 1988; Rao *et al*, 1989), and the third is the logit linear relation, which can be modeled with a logistic function (Chen, 2011). In many cases, the relationships between the level of urbanization and per capita GRP of the 31 Chinese regions satisfy a logistic function (Chen, 2011), which can be expressed as

$$L_i = \frac{L_{max}}{1+Ae^{-kG_i}}, \tag{28}$$

which can be transformed into a logarithmic linear relation, $\ln(L_{max}/L_i-1)=\ln A-kG_i$, where $L_i$ and $G_i$ denote the level of urbanization and per capita GRP of the *i*th regions, $A$, $k$, and $L_{max}$ are parameters. Among these parameters, $L_{max}$ is the capacity of the level of urbanization in a region. For simplicity, let $L_{max}$ equal 100%. A least squares calculation using the 2012's datasets consisting of 29 elements yields the following model

$$\ln(\frac{100}{L_i}-1) = 1.1201 - 0.00003022G_i + \varepsilon_i.$$

The goodness of fit is about $R^2$=0.8699, which is less to a degree than that of the linear model (Figure 3). The logarithmic linear regression can be applied to all the available datasets from 2000 to 2013 (Table 4). From 2000 to 2008, the goodness of fit of the logistic model is greater than that of the linear model, but from 2009 to 2013, the $R$ square of the linear model exceeds that of the logistic model. This suggests a complicated and evolutive correlation between urbanization and economic development.

**Table 4 The coefficients and goodness of fit of the regression models of the correlation between urbanization and economic development of 29 Chinese regions (2000-2013)**

| Model | Parameter /Statistic | 2000 | 2005 | 2006 | 2007 | 2008 | 2009 | 2010 | 2011 | 2012 | 2013 |
|---|---|---|---|---|---|---|---|---|---|---|---|

| Linear model | $a$ | 20.1216 | 24.3466 | 25.1037 | 25.7844 | 24.6789 | 25.3256 | 25.1019 | 25.1020 | 26.1393 | 27.1009 |
| --- | --- | --- | --- | --- | --- | --- | --- | --- | --- | --- | --- |
| | $b$ | 2.2724E-03 | 1.3107E-03 | 1.1510E-03 | 9.8978E-04 | 9.1474E-04 | 8.5882E-04 | 7.8448E-04 | 6.9522E-04 | 6.3884E-04 | 5.9096E-04 |
| | $R^2$ | 0.8358 | 0.8931 | 0.8925 | 0.8969 | 0.9068 | 0.9048 | 0.9172 | 0.9063 | 0.8944 | 0.8889 |
| Logistic model | $k$ | 1.0616E-04 | 6.1488E-05 | 5.3919E-05 | 4.6466E-05 | 4.2704E-05 | 4.0097E-05 | 3.6911E-05 | 3.2750E-05 | 3.0217E-05 | -2.8144E-05 |
| | $A$ | 3.8269 | 3.1806 | 3.0773 | 2.9992 | 3.1543 | 3.0745 | 3.1538 | 3.1791 | 3.0651 | 2.9713 |
| | $R^2$ | 0.8656 | 0.9126 | 0.9109 | 0.9142 | 0.9081 | 0.9002 | 0.9057 | 0.8858 | 0.8699 | 0.8611 |

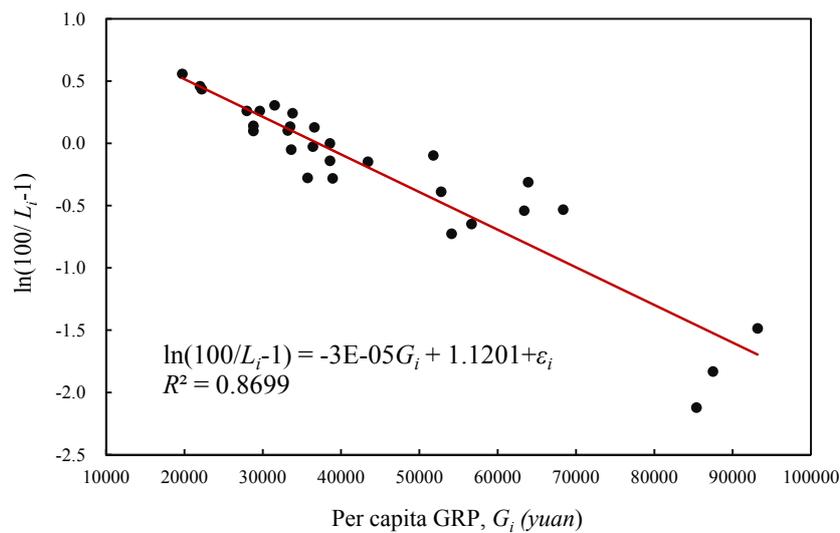

**Figure 3 The linear regression of the logistic relationship between urbanization and economic development of the 29 Chinese regions (2012)**

The method of spatial autocorrelation analysis can be applied to the residuals from the logistic models for different years. The results are tabulated as below (Table 5). The cases are similar to those of linear models (Table 3). The Durbin-Watson statistic values depend on the permutation of the 29 regions. For example, for 2000, the DW value based on the permutation in the conventional order is about 1.5870, but the result based on the alphabetical permutation is around 2.4902. There is a significant difference between the two numerical values. However, without exception, the RCI value and ARCI values are free from the influence of the arrangement order of the members in the datasets. This implies that the new approach of serial correlation test is applicative to the least squares regression based on the linearized expressions of nonlinear models.

**Table 5 The Durbin-Watson statistics, RCI values, and ARCI values of residual series from linearized logistic models of 29 Chinese regions (2000-2013)**

| Year | Arrangement in conventional order | | | | | Arrangement in alphabetical order | | | | |
|---|---|---|---|---|---|---|---|---|---|---|
| | DW statistic | Power law based | | Exponential law based | | DW statistic | Power law based | | Exponential law based | |
| | | RCI | ARCI | RCI | ARCI | | RCI | ARCI | RCI | ARCI |
| **2000** | 1.5870 | 1.7934 | 1.8068 | 1.7765 | 1.7322 | 2.4902 | 1.7934 | 1.8068 | 1.7765 | 1.7322 |
| **2005** | 1.3898 | 1.8706 | 1.8061 | 1.8782 | 1.7742 | 1.9284 | 1.8706 | 1.8061 | 1.8782 | 1.7742 |
| **2006** | 1.4574 | 1.8935 | 1.8448 | 1.9032 | 1.8215 | 1.8541 | 1.8935 | 1.8448 | 1.9032 | 1.8215 |
| **2007** | 1.5653 | 1.9246 | 1.9013 | 1.9331 | 1.8851 | 1.8928 | 1.9246 | 1.9013 | 1.9331 | 1.8851 |
| **2008** | 1.5630 | 2.0557 | 2.0749 | 2.0303 | 2.0297 | 1.9364 | 2.0557 | 2.0749 | 2.0303 | 2.0297 |
| **2009** | 1.7473 | 2.1454 | 2.1954 | 2.1034 | 2.1462 | 1.8958 | 2.1454 | 2.1954 | 2.1034 | 2.1462 |
| **2010** | 1.9178 | 2.1946 | 2.3054 | 2.1288 | 2.2451 | 1.9866 | 2.1946 | 2.3054 | 2.1288 | 2.2451 |
| **2011** | 2.0921 | 2.2599 | 2.3915 | 2.1765 | 2.3067 | 1.8872 | 2.2599 | 2.3915 | 2.1765 | 2.3067 |
| **2012** | 2.2132 | 2.2788 | 2.3946 | 2.1977 | 2.3067 | 1.8714 | 2.2788 | 2.3946 | 2.1977 | 2.3067 |
| **2013** | 2.2334 | 2.2998 | 2.4204 | 2.2219 | 2.3274 | 1.8127 | 2.2998 | 2.4204 | 2.2219 | 2.3274 |

# 4 Discussion

## 4.1 Basic framework of methodology

This paper is devoted to developing a methodology of serial correlation test for the residuals from regression models based on spatial random samples. This work is on method development rather than empirical analysis. Mathematical modeling is not the main task of this study, but the cases of regression analyses can be employed to show how to apply the spatial autocorrelation approaches to testing for serial correlation in least squares regression. Differing from the conventional Durbin-Watson statistic, the spatial DW statistics based on Moran's index and Geary's coefficient, RCI and ARCI, are independent of the permutation of elements in a dataset. This indicates that the new method is effective for testing residuals from least squares regression associated with spatial modeling. The merits of this method are as follows. First, the mathematical principles are simple and easy to understand; second, the calculation is simple and convenient to implement. Actually, we can utilize the approaches by MS Excel (S4). The processes of statistic analysis and serial correlation test based on a spatial random sample can be illustrated using a flow chart (Figure 4).

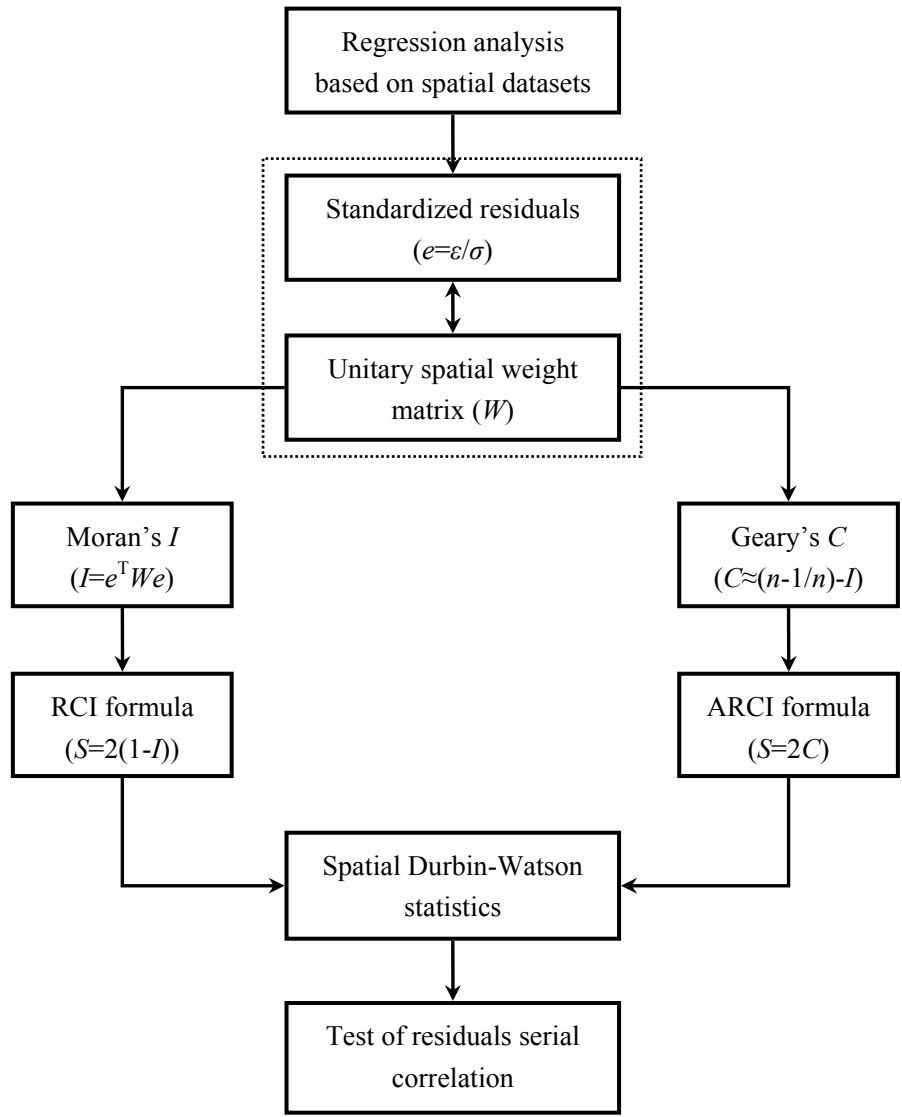

**Figure 4 A flow chart of the two spatial autocorrelation approaches to testing residuals from least squares regression based on spatial random samples**

In fact, a preliminary progress was made forty years ago, but the result failed to catch people's attention. Cliff and Ord (1973) once employed Moran's index to test the regression residuals for autocorrelation (Haggett *et al*, 1977). Compared with the previous work, the advances made in this article are as follows. First, both Moran's index and Geary's coefficient are adopted to evaluate autocorrelation of regression residuals. Cliff and Ord (1973) only made use of Moran's index. Second, two new statistics are defined by analogy with the Durbin-Watson statistic. Based on Moran's index and Geary's coefficient, the statistics termed RCI and ARCI for short are constructed. By means of RCI and ARCI, the Durbin-Watson significance tables can be utilized to form a judgment. Third, the relationships between different statistics are revealed by mathematical transformation. In this manner, it is easy to understand the statistics. Fourth, the new statistics are expressed with a matrix and a vector. The weight matrix is unitized, and the vector is standardized. So the expressions are normalized and it is convenient to compute the statistics. Fifth, typical case studies are made to demonstrate the analytical processes. According to these examples, readers can make autocorrelation tests for the regression residuals based on spatial datasets.

**4.2 Deficiency in the method**

Nothing is perfect in the world. Any measure has its shortcomings, and any method has its flaws. The incompleteness of the SDW statistic and the corresponding test method rests with SCM. In fact, the RCI values and ARCI values depend on the form of the spatial weight function. Different spatial weight functions yield different SCMs, and different SCMs result in different SDW values. In geographical analysis, we have four types of spatial weight function, including inverse power function, negative exponential function, staircase function, and semi-staircase function. The inverse power function is for the spatial processes based on globality associated with action at a distance, the negative exponential function is for those based on localization or quasi-globality, the staircase function is for those based on locality, and the semi-staircase function is for those based on quasi-locality (Chen, 2012). In many cases, it is difficult to make a decision for selection of a weight function in concrete studies. In order to choose a proper weight matrix, it is necessary to know the mathematical properties and physical meanings of different functions and the geographical features of study areas. This kind of problems is pending and remains to be solved in future studies.

# 5 Conclusions

A well-known difficult problem in spatial analysis is testing for serial correlation in least squares regression based on spatial random samples. This problem has not been solved effectively so far. In this paper, a new methodology for testing autocorrelation of residuals is illustrated, including mathematical models, statistic principles, calculation processes, and typical cases. The main conclusions can be drawn from the theoretical reasoning and empirical analyses as follows.

**First, the spatial autocorrelation analysis can be employed to test the serial correlation of residuals from least squares regression**. The formula of the Durbin-Watson statistic is a mathematical expression based on one-order time lag for ordered time series, or based on one-step spatial displacement for ordered space series. If we make a regression analysis using cross-sectional data from spatial random sampling, the Durbin-Watson test will be ineffective because the results depend on the arrangement order of elements in arrays. Rearranging the data sequences in the independent variable(s) and dependent variables will yield various DW values. In many cases, these DW values are significantly different from one another. If we use the spatial weight function to replace the parameter of time lag or space displacement, the problem of random results will be well solved for serial correlation tests.

**Second, the new statistics for testing residual correlation of spatial random series can be constructed with two related ways**. One is on the analogy of Moran's index, and the other is on the analogy of Geary's coefficient. By way of Moran's index, we can get a spatial autocorrelation coefficient of spatial residuals. One minus the coefficient is equal to half of the autocorrelation index of residual series. In other words, doubling the difference between 1 and the SAI yields the precise RCI value. By way of Geary's coefficient, we can obtain another spatial correlation index. Doubling this index gives ARCI, which is strictly equivalent to Durban-Watson statistic, and can be called SDW statistic in spatial analysis.

**Third, a theoretical expression is different from an empirical formula**. The traditional formula of Moran's index is a theoretical expression, which is based on populations. In contrast with Moran's index, Geary's coefficient is based on samples. In short, Moran's index is for theoretical derivation, while Geary's coefficient is for empirical analyses. Empirical researches are forever based on samples. Excepting the processes of mathematical reasoning, the sample size ($n$) in equations should be substituted with the total degree of freedom ($n$-1). In practice, the population standard deviation should be replaced by sample standard deviation. For convenience, the SDW statistic, i.e., ARCI, can be used to replace RCI to make a test for serial correlation in the least square regression of geographical spatial analyses, and the common Durbin-Watson significance tables can be adapted for this test of serial autocorrelation.

**Acknowledgement:**

This research was supported financially by the National Natural Science Foundation of China (Grant no. 41171129).

# References

Batty M (1991). Cities as fractals: Simulating growth and form. In: Crilly AJ, Earnshaw RA, Jones H (eds). *Fractals and Chaos*. New York: Springer–Verlag, pp43–69

Chen YG (2008). A wave-spectrum analysis of urban population density: entropy, fractal, and spatial localization. *Discrete Dynamics in Nature and Society*, vol. 2008, Article ID 728420, 22 pages

Chen YG (2011). Modelling the relationships between urbanization and economic development levels with three functions. *Scientia Geographica Sinica*, 31(1): 1-6 (In Chinese)

Chen YG (2012). On the four types of weight functions for spatial contiguity matrix. *Letters in Spatial and Resource Sciences*, 5(2): 65-72

Chen YG (2013). New approaches for calculating Moran's index of spatial autocorrelation. *PLoS ONE*, 8(7): e68336

Cliff AD, Ord JK (1973). *Spatial Autocorrelation*. London: Pion

Dillinger W (1979). A national urban data file for 114 countries. In: Urban and Regional Economics Division, World Bank (Ed). *Urban and Regional Report (Number 79-5)*. Washington, DC: World Bank

Durbin J, Watson GS (1950). Testing for serial correlation in least squares regression (I). *Biometrika*, 37 (3–4): 409–428

Durbin J, Watson GS (1951). Testing for serial correlation in least squares regression (II). *Biometrika*, 38 (1–2): 159–179

Durbin J, Watson GS (1971). Testing for serial correlation in least squares regression (III). *Biometrika*, 58 (1): 1–19

Fotheringham AS, O'Kelly ME (1989). *Spatial Interaction Models: Formulations and Applications*. Boston: Kluwer Academic Publishers

Geary RC (1954). The contiguity ratio and statistical mapping. *The Incorporated Statistician*, 5(3): 115–145

Haggett P, Cliff AD, Frey A (1977). *Locational Analysis in Human Geography*. London: Edward Arnold

Kac M (1969). Some mathematical models in science. *Science*, 166: 695-699

Longley PA (1999). Computer simulation and modeling of urban structure and development. In: M. Pacione (Ed). *Applied Geography: Principles and Practice*. London and New York: Routledge, pp605-619

Moran PAP (1950). Notes on continuous stochastic phenomena. *Biometrika*, 37(1-2): 17-33

Rao DN, Karmeshu, Jain VP (1988). Model of urbanization based on replacement dynamics—an international comparison of empirical estimates. In: *Natural Workshop on Futurology, Technological Forecasting towards the 21st Century*. New Delhi, India: Indian Institute of Technology

Rao DN, Karmeshu, Jain VP (1989). Dynamics of urbanization: the empirical validation of the replacement hypothesis. *Environment and Planning B: Planning and Design*, 16(3): 289-295

Sargan JD, Bhargava A (1983). Testing residuals from least squares regression for being generated by the Gaussian random walk. *Econometrica*, 51(1): 153-174

Zhou YX (1989). On the relationship between urbanization and gross national product. *Chinese Sociology & Anthropology*, 21(2): 3-16

# Appendices

## Appendix 1: Pearson's correlation coefficient and Moran's index

The mathematical expression of Moran's index can be derived by generalizing the formula of Pearson's correlation coefficient step by step. First, generalize the zero-order cross-correlation coefficient (time lag is 0) to a one-order autocorrelation coefficient defined in a 1-dimension time (time lag is 1). Second, generalize one-order temporal autocorrelation coefficient to a one-order autocorrelation coefficient defined in a 1-dimension space (the time lag is replaced by spatial displacement, and the displacement is 1). Third, generalize one-order spatial autocorrelation coefficient to a weighted autocorrelation coefficient defined in a 2-dimension space (the spatial displacement is replaced by spatial weight). The result is the 2-dimension spatial autocorrelation coefficient termed Moran's index.

The coefficient of Pearson's correlation is actually a zero-order cross-correlation coefficient. That is, the time lag or spatial displacement of correlation is 0. Suppose there is random sample including $n$ elements. The sample can be measured with two variables, $x$ and $y$. The correlation coefficient can be expressed as

$$R = \frac{\sum_{i=1}^{n}(x_i - \bar{x})(y_i - \bar{y})}{\sqrt{\sum_{i=1}^{n}(x_i - \bar{x})^2 \sum_{i=1}^{n}(y_i - \bar{y})^2}}, \tag{A1}$$

where $R$ denotes correlation coefficient, $i=1,2,\ldots,n$, and

$$\bar{x} = \frac{1}{n}\sum_{i=1}^{n} x_i, \quad \bar{y} = \frac{1}{n}\sum_{i=1}^{n} y_i$$

are two average values of the random variables, $x$ and $y$. Equation (A1) represents the simplest cross-correlation coefficient with time lag is $\tau=0$.

If the sample is described with only one variable, $x$, then Pearson's correlation coefficient can be generalized to one-order autocorrelation coefficient as below

$$R = \frac{\sum_{t=1}^{n-1}(x_t - \bar{x}_{1,n-1})(x_{t+1} - \bar{x}_{2,n})}{\sqrt{\sum_{t=1}^{n-1}(x_t - \bar{x}_{1,n-1})^2 \sum_{t=1}^{n-1}(x_{t+1} - \bar{x}_{2,n})}}, \tag{A2}$$

where $t$ represents time, and

$$\bar{x}_{1,n-1} = \frac{1}{n-1}\sum_{t=1}^{n-1} x_t, \quad \bar{x}_{2,n} = \frac{1}{n-1}\sum_{t=1}^{n-1} x_{t+1}$$

refer to average values. For equation (A2), the time lag is $\tau=1$. Apparently, equations (A1) and (A2) are identical in mathematical structure to each other. Equation (A1) reflects the cross-correlation between two random variables, while equation (A2) reflects the autocorrelation of a variable. For simplicity, equations (A2) can be approximately revised as

$$R = \frac{\sum_{t=1}^{n-1}(x_t - \bar{x})(x_{t+1} - \bar{x})}{\sqrt{\sum_{t=1}^{n-1}(x_t - \bar{x})^2 \sum_{t=1}^{n-1}(x_{t+1} - \bar{x})^2}}, \tag{A3}$$

in which the mean value is

$$\bar{x} = \frac{1}{n}\sum_{t=1}^{n} x_t .$$

If the sample size $n$ is large enough, we will have

$$\sum_{t=1}^{n-1}(x_t - \bar{x})^2 \approx \sum_{t=1}^{n-1}(x_{t+1} - \bar{x}) \approx \sum_{t=1}^{n}(x_t - \bar{x})^2 . \tag{A4}$$

Then the autocorrelation coefficient can be further revised as

$$R = \frac{\sum_{t=2}^{n}(x_t - \bar{x})(x_{t-1} - \bar{x})}{\sum_{t=1}^{n}(x_t - \bar{x})^2} , \tag{A5}$$

which is the standard expression of autocorrelation coefficient defined in 1-dimension time. Equation (A2) has small variability but large bias, while equation (A5) has large variability but small bias. Variability means deviation. The negative effect of bias is bigger than variability in statistic analysis. Therefore we choose equation (A5) over equation (A2). Substituting a distance variable, $r$, for the time variable, $t$, in equation (A5) yields

$$R = \frac{\sum_{r=2}^{n}(x_r - \bar{x})(x_{r-1} - \bar{x})}{\sum_{r=1}^{n}(x_r - \bar{x})^2} , \tag{A6}$$

which is the zero-order spatial autocorrelation coefficient defined in a 1-dimension space. The spatial displacement is $\rho=1$.

The 1-dimension spatial autocorrelation coefficient can be generalized to 2-dimension spatial autocorrelation coefficient. Consider $n$ elements in a geographical region. The elements can be arranged in various ways in a spreadsheet. For each element, the spatial autocorrelation coefficient can be expressed as

$$R_j = \frac{\sum_{i=1}^{n-1}(x_i - \bar{x})(x_{i+1} - \bar{x})}{\sum_{i=1}^{n}(x_i - \bar{x})^2} , \tag{A7}$$

where $j=1,2,\ldots,n$. For the $n$ elements, the mean of the spatial autocorrelation coefficients is

$$I = \frac{1}{n}\sum_{j=1}^{n} R_j = \frac{\sum_{j=1}^{n}\sum_{i=1}^{n-1}(x_i - \bar{x})(x_{i+1} - \bar{x})}{n\sum_{i=1}^{n}(x_i - \bar{x})^2}, \quad (A8)$$

where $I$ is an average measurement. Given an even spatial weight $w_{ij}=1/n^2$, it follows

$$I = \frac{n^2 \sum_{i=1}^{n}\sum_{j=1}^{n}\frac{1}{n^2}(x_i - \bar{x})(x_j - \bar{x})}{n\sum_{i=1}^{n}(x_i - \bar{x})^2} \Rightarrow \frac{n\sum_{i=1}^{n}\sum_{j=1}^{n}w_{ij}(x_i - \bar{x})(x_j - \bar{x})}{\sum_{i=1}^{n}(x_i - \bar{x})^2}, \quad (A9)$$

in which the weight $w_{ij}$ meets the following condition

$$\sum_{i=1}^{n}\sum_{j=1}^{n} w_{ij} = 1. \quad (A10)$$

However, in practice, the abovementioned condition cannot be satisfied. Let

$$W_0 = \sum_{i=1}^{n}\sum_{j=1}^{n} w_{ij}. \quad (A11)$$

Then equation (A9) can be revised as

$$I = \frac{\sum_{i=1}^{n}\sum_{j=1}^{n}\frac{w_{ij}}{W_0}(x_i - \bar{x})(x_j - \bar{x})}{\frac{1}{n}\sum_{i=1}^{n}(x_i - \bar{x})^2} = \frac{1}{S^2}\sum_{i=1}^{n}\sum_{j=1}^{n}\frac{w_{ij}}{W_0}(x_i - \bar{x})(x_j - \bar{x}), \quad (A12)$$

which is just the formula of Moran's index. In equation (A12), the spatial displacement parameter is substituted with a spatial weight function.

## Appendix 2: Durbin-Watson's statistic and Geary's coefficient

Derivation of Geary's coefficient is involved with several mathematical processes such as analogy, deduction, and generalization. This differs from the proof of a theorem. Suppose there are $n$ elements in a geographical region. The formula of Durbin-Watson's statistic of a linear regression model can be expressed as follows

$$DW = \frac{\sum_{i=2}^{n}(e_i - e_{i-1})^2}{\sum_{i=1}^{n}e_i^2} = \frac{\sum_{i=2}^{n}[(y_i - \hat{y}_i) - (y_{i-1} - \hat{y}_{i-1})]^2}{\sum_{i=1}^{n}(y_i - \hat{y}_i)^2} = \frac{\sum_{i=2}^{n}[(e_i - \bar{e}) - (e_{i-1} - \bar{e})]^2}{\sum_{i=1}^{n}(e_i - \bar{e})^2}. \quad (B1)$$

in which $i=1,2,\ldots,n$, and

$$e_i = y_i - \hat{y}_i \tag{B2}$$

denotes the residuals of a regression model prediction, $y_i$ refer to observed value of the dependent variable, and $y_i$ hat to the corresponding predicted value.

In order to generalize the Durbin-Watson statistic to Geary's coefficient, we must draw an analogy between similar variables. First, the deviation of the independent variable, $x_i - \bar{x}$, bears an analogy with the error, $y_i - \hat{y}_i$, where $\bar{x}$ is the mean of $x$. Second, the deviation, $x_i - \bar{x}$, have an analogy with the deviation of the error, $e_i - \bar{e}$, where $\bar{e} = 0$ indicates the mean of $e$. For any $x_i$, the Durbin-Watson statistic can be generalized from the time domain to the space domain, and we have

$$DW_j = \frac{\sum_{i=1}^{n}[(x_i - \bar{x}) - (x_j - \bar{x})]^2}{\sum_{i=1}^{n}(x_i - \bar{x})^2} = \frac{\sum_{i=1}^{n}(x_i - x_j)^2}{\sum_{i=1}^{n}(x_i - \bar{x})^2}, \tag{B3}$$

which is defined in a 1-dimension time or space.

Further, we can generalize the Durbin-Watson statistic from the 1-dimension time/space to a 2-dimension space. Taking the mean of the DW values based on $n$ points yields

$$\overline{DW} = \frac{1}{n}\sum_{i=j}^{n}DW_j = \frac{1}{n}\frac{\sum_{j=1}^{n}\sum_{i=1}^{n}(x_i - x_j)^2}{\sum_{i=1}^{n}(x_i - \bar{x})^2} = \frac{\sum_{i=1}^{n}\sum_{j=1}^{n}(x_i - x_j)^2}{n\sum_{i=1}^{n}(x_i - \bar{x})^2}. \tag{B4}$$

Because of the symmetry of the spatial contiguity matrix, only the upper triangular matrix or the lower triangular matrix will be considered in our spatial analysis. The number of spatial correlation based on the lower/upper triangular matrix is as below

$$S = 1 + 2 + \cdots + n = \frac{1}{2}n(n+1). \tag{B5}$$

The diagonal elements in the spatial contiguity matrix reflect the zero-order autocorrelation of the geographical elements and give no essential spatial information. Therefore, the effective number of spatial correlation based on the lower/upper triangular matrix is

$$S - n = 1 + 2 + \cdots + n - 1 = \frac{1}{2}(n-1)n. \tag{B6}$$

Introducing an even spatial weight based on equation (B6), $1/(S-n)$, into equation (B4) yields

$$C = \frac{\frac{1}{2}(n-1)n\sum_{i=1}^{n}\sum_{j=1}^{n}\frac{1}{S-n}(x_i-x_j)^2}{n\sum_{i=1}^{n}(x_i-\bar{x})^2} = \frac{(n-1)\sum_{i=1}^{n}\sum_{j=1}^{n}\frac{1}{S-n}(x_i-x_j)^2}{2\sum_{i=1}^{n}(x_i-\bar{x})^2}, \quad (B7)$$

in which the symbol $DW$ has been substituted with a correlation measurement $C$.

However, the geographical space in the real world is of heterogeneity instead of homogeneity, the even weight function $1/(S\text{-}n)$ should be replaced by an uneven weight such as

$$W = \frac{w_{ij}}{\sum_{i=1}^{n}\sum_{j=1}^{n}w_{ij}}. \quad (B8)$$

Thus equation (B7) can be revised as

$$C = \frac{(n-1)\sum_{i=1}^{n}\sum_{j=1}^{n}w_{ij}(x_i-x_j)^2}{2W_0\sum_{i=1}^{n}(x_i-\bar{x})^2} = \frac{(n-1)\sum_{i=1}^{n}\sum_{j=1}^{n}w_{ij}(x_i-x_j)^2}{2\sum_{i=1}^{n}\sum_{j=1}^{n}w_{ij}\sum_{i=1}^{n}(x_i-\bar{x})^2}, \quad (B9)$$

where

$$W_0 = \sum_{i=1}^{n}\sum_{j=1}^{n}w_{ij}, \quad (B10)$$

in which the spatial weight $w_{ij}$ can be generated by a power-law deccay function, an exponential decay function, or unit-step function. Equation (B9) is just the formula of Geary's coefficient.